\documentstyle[12pt]{article}
\begin{document}
\begin{titlepage}
\begin{flushright}
December, 1995
\end{flushright}
\vspace{3cm}
\begin{center}
{ {\bf \LARGE CP violation in Hyperon decays}}\\
\vspace{1cm}
{Darwin Chang and Chuan-Hung Chen}\\
\vspace{1cm}
{\Large {\em Department of physics}}\\
{\Large {\em National Tsing Hua University}}\\
{\Large {\em Hsinchu, Taiwan 30043, R.O.C.}}\\
\vspace{5cm}
{\bf Abstract}\\
\end{center}
The CP properties in hyperon decays are briefly reviewed.
We discuss the general phenomenology and define CP odd observables in 
hyperon decays.  With these observables, we discuss the predictions 
of some models and their observational potential.

\end{titlepage}
\baselineskip=19pt

\section {Introduction}
\ \ \ \ 
More than thirty years after its discovery\cite{CCFT}, the origin of CP 
violation still puzzles most of theoretical and experimental physicists. 
It remains one of the most important questions in physics. 
Until today, the effects of CP violation has been measured only in kaon 
system. In kaon deacys, two parameters, $\epsilon$ and $\epsilon'$, 
are needed to describe CP violation which represent the indirect 
and direct CP violating mechanism.  The current experimental value for 
$\epsilon$ is\cite{PDG} 
\begin{equation}
|\epsilon|=2.26 \times 10^{-3},
\end{equation}
In suitable convention, $\epsilon$ can be associated with the imaginary 
part of $K-\bar{K}$ mixing; and it denotes $|\Delta S|=2$ process.
However, the experimental situation for direct CP violating parameter 
$\epsilon'$ is still ambiguous.  Its measured values are \cite{NA31,E731}
\begin{eqnarray}
\frac{\epsilon'}{\epsilon}\:=\:\cases {(2.3\pm0.65)\times
10^{-3} & NA31, \cr (0.74 \pm 0.52 \pm 0.29)\times 10^{-3} & E731.\cr}
\end{eqnarray}
If $\epsilon'$ turns out to be nonzero, it would indicate the existence of
direct $|\Delta S|=1$ CP violation.  
Nonleptonic hyperon decays of $\Lambda$, $\Sigma$ and $\Xi$ 
provide a direct probe into the potential $\Delta S=1$ CP nonconservation. 
It is a particularly interesting system because it contains more than one 
angular momentum channels and more than one isospin channels also.  
Therefore the system provides a variety of ways for amplitudes to 
interefere which is essential for CP violation to be observable. 
It is also interesting because E871 experiment at Fermilab \cite{GHL} 
will measure CP violation in $\Xi^{-} \rightarrow \Lambda \pi^{-}$ and
$\Lambda \rightarrow p \pi^{-}$ using $2 \time 10^9$ $\Xi^-$ and 
$\bar{\Xi}^+$ produced by colliding 800 GeV proton with fixed target.  
$\Xi \rightarrow \Lambda$ is particularly interesting because one needs spin 
information to measure CP violation in hyperon decays.  The spin 
information for $\Lambda$ can be obtained from the angular distribution 
of its two body decay.  
Experience from E756 indicates they can collect $2 \times 10^7$ events 
per day with very simple trigger.
With 200 days running time, the expected sensitivity for CP
violating asymmetry is about $10^{-4}$ for A($\Lambda^{0}_{-}$)
+ A($\Xi^{-}_{-}$) in the first run which is not too far from the 
prediction of SM of a few $\times 10^{-5}$ with large uncertainty.  If 
The sensitivity may be improved in the future runs.  The main limiting 
factor seems to be whether the chamber can tolerate the large flux 
corresponding to collecting 1400 $\bar{\Xi}^+$ decays per second.
Theoretically the 
most serious obstacle against having large CP odd observable, A, comes 
from the serious constraint from the current limit on 
$\epsilon'$.  However, the two parameters in principle measure different 
$\Delta S = 1$ interactions.  One of the theoretical challenges is to 
disintangle the two obersvables within various models.

This article is organized as follows. In section 2 we make the 
phenomenological analysis, define the CP odd observables and formulate 
the relationships between the phenomenological parameters and the CP odd 
observables. In section 3 we compare the predictions of various models, 
including SM.  Finally we give some concluding remarks.

\section {Phenomenological Analysis}

\subsection {CP odd observables}
\ \ \ \ 
For $J^{P}={\frac{1}{2}}^{+}$ hyperon, the most general decay amplitude 
is \cite{TDLEE,COMMINS,Marshark}
\begin{eqnarray}
{\cal M} = G_{F}m^{2}_{\pi}\bar{u_{f}}(a-b\gamma_{5})u_{i}, 
\end {eqnarray} 
where a and b are constants. $G_{F}$ is the Fermi constant,
$m_{\pi}$ is the mass of $\pi$ meson and $u_{i,f}$
denote the initial and final baryon states. Due to the negative intrinsic
parity of $\pi$ meson, the a and b terms denote parity violating
and conserving process respectively. 
The matrix element can be reduced to 
\begin{equation}
{\cal M}(B_{i} \rightarrow B_{f} \pi)=S+P\vec{\sigma} \cdot \hat{q},
\end{equation}
where $\hat{q}$ is the direction of final baryon $B_{f}$ momentum. $S= a$ is 
the amplitude with the final state in the S-wave, $\ell=0$, parity-odd 
state; and $P = |\vec{q}| b /(E_f + M_f)$ denotes the amplitude with the 
final state in the P-wave, $\ell=1$, parity-even state. 
In terms of these quantities the transition rate can be denoted by
\begin{eqnarray}
\frac{4\pi}{\Gamma}\frac{d\Gamma}{d\Omega}=1+\alpha\hat{s}_{i}\cdot\hat{q}
+ \hat{s}_{f}\cdot[(\alpha+\hat{s}_{i}\cdot\hat{q})\hat{q}+\beta\hat{s}_{i}
\times \hat{q}+\gamma(\hat{q}\times (\hat{s}_{i}\times \hat{q}))],
\end{eqnarray}
in the rest frame of the initial hyperon, where $\hat{s}_{i,f}$ are
the spin vectors of the initial and final baryons, respectively. 
$\Gamma$ is the total decay rate given by
\begin{equation}
\Gamma=\frac{|\vec{q}|(E_{f}+E_{i})}{4\pi m_{i}}G^{2}_{F}m^{4}_{\pi}
(|S|^{2}+|P|^{2}).\label{eqn:ga}
\end{equation}
The parameters of $\alpha$, $\beta$ and $\gamma$ is defined as
\begin{eqnarray}
\alpha=2\frac{Re(S^{*}P)}{|S|^{2}+|P|^{2}},\:\: \beta=2\frac{Im(S^{*}P)}{
|S|^{2}+|P|^{2}},\:\: \gamma=\frac{|S|^{2}-|P|^{2}}{|S|^{2}+|P|^{2}}.
\end{eqnarray}
The three parameters, $\alpha$, $\beta$ and $\gamma$, are not independent, 
since 
\begin{equation}
\alpha^{2}+\beta^{2}+\gamma^{2}=1.
\end{equation}
Therefore $\beta$, $\gamma$ can be parametrized by an angle $\phi$ as 
$\beta=(1-\alpha^{2})^{1/2}sin\phi$, $\gamma=(1-\alpha^{2})^{1/2}cos\phi$.
Note that $\alpha$ 
involves only one spin and it is $\hat{T}$ even 
(we shall define $\hat{T}$ as the transformation that reverse 
all the directions of all the vectors but keep the initial state and 
final state intact); 
$\beta$ involves only two spins and is $\hat{T}$ odd and 
$\gamma$ involves only two spins and is $\hat{T}$ even.  Therefore, 
$\beta$ and $\gamma$, which involves two and three vectors respectively,
are intrinsincally harder to measure than $\alpha$.
Since $\hat{T}$ is not time reversal, even without CP violation, the 
final state interaction (FSI) can produce $\hat{T}$-odd effect also.
Therefore, a nonvanishing $\beta$ can be due to either FSI or CP violation. 

If one does not measure $\hat{s}_f$, the decay distribution of $B_f$ is 
\begin{equation}
{4 \pi \over \Gamma} \frac{d\Gamma}{d\Omega}=1+\alpha \vec{P}_i \cdot\hat{q},
\end{equation}
where $\vec{P}_i$ is the polarization of $B_i$.  Therefore, the decay 
distribution of $B_f$ can be used to measure the polarization of $B_i$.
In general, the polarization of $B_f$ is given by 
\begin{eqnarray}
\vec{P}_f = \frac{(\alpha+\vec{P}_{i}\cdot\hat{q})\hat{q}+\beta\vec{P}_{i}
\times \hat{q}+\gamma(\hat{q}\times (\vec{P}_{i}\times \hat{q}))}
{1 + \alpha\vec{P}_{i}\cdot\hat{q}}.
\end{eqnarray}
Therefore, $\vec{P}_f = \alpha \hat{q}$ if $\vec{P}_i =0$.  In $\Xi^- 
\rightarrow \Lambda + \pi \rightarrow p \pi \pi$, if $\Xi^-$ is produced 
unpolarized as in E871 experimental design, then $\vec{P}_{\Lambda} = 
\alpha_{\Xi} \hat{p}_{\Lambda}$, and the decay distribution proton is 
\begin{equation}
{4\pi \over \Gamma_p} \frac{d\Gamma_p}{d\Omega} = 1+\alpha_{\Lambda}
\vec{P}_{\Lambda} \cdot\hat{q}
= 1+ {\bf a} \hat{p}_{\Lambda} \cdot\hat{q},
\end{equation}
where ${\bf a}$ is $\alpha_{\Lambda} \alpha_{\Xi}$.  Similarly one can define
$\bar{{\bf a}}$ in $\bar{\Xi}$ decay.  E871 experiment will measure $\delta {\bf a}
=({\bf a}-\bar{{\bf a}})/({\bf a}+\bar{{\bf a}})$ which is rouhgly
$A(\Lambda) + A(\Xi)$.  Note
however, there may be a small polarization of $\Xi$ due to the 
interference between strong and weak process in the production.  In that 
case, one can still measure $\Xi$ polarization directly by determining the 
$\Lambda$ momentum precisely.  

The amplitudes of S and P in Eq.(4), in general, are complex
numbers.  They can be parametrized as\cite{WC,DP}:
\begin{eqnarray}
S=\sum_{i}S_{i}e^{i(\delta^{S}_{i}+\theta^{S}_{i})},\nonumber \\
P=\sum_{i}P_{i}e^{i(\delta^{P}_{i}+\theta^{P}_{i})}.\label{eqn:sp}
\end{eqnarray}
Here we have extracted the strong rescattering phases
$\delta_{i}$ and the weak CP violating phases $\theta_{i}$ from the 
amplitudes and 
$S_{i}$ and $P_{i}$ are real where $i$ represents all possible final 
isospin states with changes in isospin $\Delta I$.
The parameters for antihyperon decays can be defined similarly, but 
denoted by $\bar{\Gamma}$, $\bar{\alpha}$, $\bar{\beta}$ and
$\bar{\gamma}$. 
In our convention, the antihyperon decay amplitudes are
\begin{eqnarray}
\bar{S}&=&-\sum_{i}S_{i}e^{i(\delta^{S}_{i}-\theta^{S}_{i})},
\nonumber \\
\bar{P}&=&\sum_{i}P_{i}e^{i(\delta^{P}_{i}-\theta^{P}_{i})}.
\label{eqn:barsp}
\end{eqnarray} 
The minus($-$) sign in the S-wave amplitude comes from the fact
that pion is parity-odd, that is,
under parity transformation parity violating $S$ amplitude will appear 
with an extra $-$ sign in the amplitude. On the other hand, 
according to CPT theorem, the amplitude of hyperon in weak interaction
will relate to the amplitude of antihyperon by complex conjugation. 
So, the amplitudes of antihyperon
are shown as Eq.(\ref{eqn:barsp}). Using Eq.(\ref{eqn:sp})
and Eq.(\ref{eqn:barsp}) we can show
\begin{eqnarray}
\alpha \stackrel{CP} {\longleftrightarrow} -\bar{\alpha},\nonumber \\
\beta \stackrel{CP} {\longleftrightarrow} -\bar{\beta}.\label{eqn:ab}
\end{eqnarray}
Using these parameters, we can define some CP asymmetric quantities in 
hyperon decays \cite{DP}: 
\begin{eqnarray}
\Delta&\equiv& \frac{\Gamma-\bar{\Gamma}}{\Gamma+\bar{\Gamma}}, \nonumber \\
    A &\equiv& \frac{\Gamma \alpha + \bar{\Gamma}\bar{\alpha}}
               {\Gamma \alpha - \bar{\Gamma}\bar{\alpha}}, \nonumber \\
    B &\equiv& \frac{\Gamma \beta + \bar{\Gamma}\bar{\beta}}
               {\Gamma \beta - \bar{\Gamma}\bar{\beta}}, \nonumber \\
    B' &\equiv& \frac{\Gamma \beta + \bar{\Gamma}\bar{\beta}}
               {\Gamma \alpha - \bar{\Gamma}\bar{\alpha}}. \label{eqn:dd}
\end{eqnarray} 
According to the results of Eq.(\ref{eqn:ab}) and Eq.(\ref{eqn:dd}),
when CP is conserved we have $\Gamma = \bar{\Gamma}$
$\alpha=-\bar{\alpha}$ and $\beta=-\bar{\beta}$; and CP
asymmetric quantities, defined in Eq.(\ref{eqn:dd}), all vanish.
When CP$\hat{T}$ is conserved, we have $\Gamma = \bar{\Gamma}$
$\alpha=-\bar{\alpha}$ and $\beta= + \bar{\beta}$.  Therefore, to have 
nonzero $\Delta$, one needs FSI and interference between different 
isospin channels (instead of momentum channels).  Therefore $\Delta$ is 
necessarily suppressed by $\Delta I = 1/2$ rule.  On the other hand, again 
due to the $\Delta I =1/2$ rule, the leading contributions to $A$, $B$ 
and $B'$  
should be due to the interference between different momentum channels(S 
and P waves channels).  Nonvanishing $A$ also requires FSI.  On the other 
hand $B$ actually gets a spurious enhancement due to the FSI suppression 
in the denominator.
In the following section we will discuss the simplified relation between
CP violating phases and physical measurements.

\subsection {Isospin Decomposition}
\ \ \ \
The Hamiltonian of $\Delta S=1$ in hyperon decays includes two different
isospin terms that one is $\Delta I=1/2$ and 
$\Delta I=3/2$ is another one. Therefore, 
using the parametrization of Eq.(\ref{eqn:sp}) we can write them to be
\cite{OP}
\begin{eqnarray}
\Lambda \longrightarrow p\pi^{-}:\ \ &&S(\Lambda^{0}_{-})=
-\sqrt{\frac 23}S_{11}e^{i(\delta_{1}+\theta^{S}_{1})} +
\sqrt{\frac 13}S_{33}e^{i(\delta_{3}+\theta^{S}_{3})},\nonumber \\
&&P(\Lambda^{0}_{-})=
-\sqrt{\frac 23}P_{11}e^{i(\delta_{11}+\theta^{P}_{1})} +
\sqrt{\frac 13}P_{33}e^{i(\delta_{31}+\theta^{P}_{3})},\nonumber \\
\Lambda \longrightarrow n\pi^{0}:\ \ &&S(\Lambda^{0}_{0})=
\sqrt{\frac 13}S_{11}e^{i(\delta_{1}+\theta^{S}_{1})} +
\sqrt{\frac 23}S_{33}e^{i(\delta_{3}+\theta^{S}_{3})},\nonumber \\
&&P(\Lambda^{0}_{0})=
\sqrt{\frac 13}P_{11}e^{i(\delta_{11}+\theta^{P}_{1})} +
\sqrt{\frac 23}P_{33}e^{i(\delta_{31}+\theta^{P}_{3})},\nonumber \\
\Xi^{-} \longrightarrow \Lambda\pi^{-}:\ \ &&S(\Xi^{-}_{-})=
S_{12}e^{i(\delta_{2}+\theta^{S}_{12})}+{\frac 12} S_{32}
e^{i(\delta_{2}+\theta^{S}_{32})}, \nonumber \\
&&P(\Xi^{-}_{-})=
P_{12}e^{i(\delta_{21}+\theta^{P}_{12})}+{\frac 12} P_{32}
e^{i(\delta_{21}+\theta^{P}_{32})}, \nonumber \\
\Xi^{0} \longrightarrow \Lambda\pi^{0}:\ \ &&S(\Xi^{0}_{0})=
\sqrt{\frac 12} (S_{12}e^{i(\delta_{2}+\theta^{S}_{12})}-
S_{32}e^{i(\delta_{2}+\theta^{S}_{32})}), \nonumber \\
&&P(\Xi^{0}_{0})=
\sqrt{\frac 12}(P_{12}e^{i(\delta_{21}+\theta^{P}_{12})}- P_{32}
e^{i(\delta_{21}+\theta^{P}_{32})}).
\end{eqnarray}
Where $S_{ij}, P_{ij}$ correspond to $S_{2\Delta I,2I}, P_{2\Delta I,2I}$,
and $\delta_{2I}$ and $\delta_{2I,1}$ for S-wave and P-wave amplitudes
, respectively. It is well known experimentally
that the $\Delta I=1/2$ amplitude is dominant. If we take first order 
in the $\Delta I=3/2$ amplitudes, we can simplify Eq.(13) as follows
\cite{DP}:
{
\setcounter{enumi}{\value{equation}}
\addtocounter{enumi}{1}
\setcounter{equation}{0}
\renewcommand{\theequation}{\theenumi.\alph{equation}}
\begin{eqnarray}
\Delta(\Lambda^{0}_{-})&=& \sqrt{2}\frac{S_{33}}{S_{11}}
sin(\delta_{3}-\delta_{1})sin(\theta^{S}_{3}-\theta^{S}_{1}),
\label{eqn:DL31}\\
A(\Lambda^{0}_{-})&=&-tan(\delta_{11}-\delta_{1})sin(\theta^{P}_{1}-
\theta^{S}_{1}) \times \nonumber \\
 &&\left[ 1+\frac{1}{\sqrt{2}} \frac{S_{33}}{S_{11}}
\left[ {cos(\delta_{11}-\delta_{3}) \over cos(\delta_{11}-\delta_{1})}
- {sin(\delta_{11}-\delta_{3}) \over sin(\delta_{11}-\delta_{1})} 
{sin(\theta^{P}_{1}-\theta^{S}_{3}) \over 
sin(\theta^{P}_{1}-\theta^{S}_{1})}
\right]\right. \nonumber \\  
 &&\left.+\frac{1}{\sqrt{2}} \frac{P_{33}}{P_{11}}
\left[{cos(\delta_{31}-
\delta_{1}) \over cos(\delta_{11}-\delta_{1})}-{sin(\delta_{31}-\delta_{1})
\over sin(\delta_{11}-\delta_{1})}
{sin(\theta^{P}_{3}-\theta^{S}_{1}) \over sin(\theta^{P}_{1}-
\theta^{S}_{1})}
\right] \right],\label{eqn:AL111}\\
B(\Lambda^{0}_{-})&=&cot(\delta_{11}-\delta_{1})sin(\theta^{P}_{1}-
\theta^{S}_{1}) \times \nonumber \\
 &&\left[ 1+\frac{1}{\sqrt{2}} \frac{S_{33}}{S_{11}}
\left[ {sin(\delta_{11}-\delta_{3}) \over sin(\delta_{11}-\delta_{1})}
- {cos(\delta_{11}-\delta_{3}) \over cos(\delta_{11}-\delta_{1})}
{sin(\theta^{P}_{1}-\theta^{S}_{3}) \over 
sin(\theta^{P}_{1}-\theta^{S}_{1})}
\right]\right. \nonumber \\  
 &&\left.+\frac{1}{\sqrt{2}} \frac{P_{33}}{P_{11}}
\left[{sin(\delta_{31}-
\delta_{1}) \over sin(\delta_{11}-\delta_{1})}-{cos(\delta_{31}-\delta_{1})
\over cos(\delta_{11}-\delta_{1})}
{sin(\theta^{P}_{3}-\theta^{S}_{1}) \over sin(\theta^{P}_{1}-
\theta^{S}_{1})}
\right] \right].\label{eqn:BL111}
\end{eqnarray}
\setcounter{equation}{\value{enumi}}
}
To first order in the $\Delta I=3/2$ amplitudes we have
\begin{eqnarray}
\Delta(\Lambda^{0}_{0})&=&-\frac 12 \Delta(\Lambda^{0}_{-}),\nonumber\\
A(\Lambda^{0}_{0})&=&A(\Lambda^{0}_{-}),\nonumber\\
B(\Lambda^{0}_{0})&=&B(\Lambda^{0}_{-}).
\end{eqnarray}
From N$\pi$ scattering, the strong rescattering phases for
 $\Lambda$ decay can be determined to be \cite{Roper}
\begin{eqnarray}
\delta_{1}\approx6.0^{o},\ \delta_{3}\approx-3.8^{o},\nonumber \\
\delta_{11}\approx-1.1^{o},\ \delta_{31}\approx-0.7^{o} \label{eqn:Lps}
\end{eqnarray}
with errors of order $1^{o}$. 
The amplitudes giving
Experimental measurements that reflect $\Delta I=1/2$ rule give
$S_{33}/S_{11}=0.027\pm0.008$ and $P_{33}/P_{11}=0.03\pm0.037$
\cite{Overseth}.

The decay amplitudes for $\Xi^{-}\rightarrow \Lambda^{0}\pi^{-}$ are
{
\setcounter{enumi}{\value{equation}}
\addtocounter{enumi}{1}
\setcounter{equation}{0}
\renewcommand{\theequation}{\theenumi.\alph{equation}}
\begin{eqnarray}
\Delta(\Xi^{-}_{-})&=&0,\\
A(\Xi^{-}_{-})&=&- tan(\delta_{21}-\delta_{2})
\left[ sin(\theta^{P}_{12}-\theta^{S}_{12}) \right. \nonumber \\
 &&\left.+\frac 12 \frac{P_{32}}{P_{12}}
(sin(\theta^{S}_{32}-\theta^{P}_{12})-1)
+\frac 12 \frac{S_{32}}{S_{12}}
(sin(\theta^{P}_{12}-\theta^{S}_{32})-1)\right],\\
B(\Xi^{-}_{-})&=&cot(\delta_{21}-\delta_{2})
\left[ sin(\theta^{P}_{12}-\theta^{S}_{12}) \right. \nonumber\\
 &&\left.+\frac 12 \frac{P_{32}}{P_{12}}
(sin(\theta^{S}_{32}-\theta^{P}_{12})-1) 
+\frac 12 \frac{S_{32}}{S_{12}}
(sin(\theta^{P}_{12}-\theta^{S}_{32})-1)\right].
\end{eqnarray}
\setcounter{equation}{\value{enumi}}
}
If $\Delta I=3/2$ contributions are treated to lowest order, we have
\begin{eqnarray}
\Delta(\Xi^{0}_{0})&=&0,\nonumber\\
A(\Xi^{0}_{0})&=&A(\Xi^{-}_{-}),\nonumber  \\ 
B(\Xi^{0}_{0})&=&B(\Xi^{-}_{-}).
\end{eqnarray}
The strong rescattering phases for $\Xi$ decays hasnot been measured 
experimentally.
Recently, Lu, Savage and Wise, 
using chiral perturbation theory, predict $\delta_{21}=-1.7^{0}$
and $\delta_{2}=0$ \cite{LSW}. The ratio of $\Delta I=3/2$ amplitudes to
that of $\Delta I=1/2$ in $\Xi$ decays are similar to 
those in $\Lambda$ decays and the experimental measurements are:
$S_{32}/S_{12}=-0.046\pm0.014$ and $P_{32}/P_{12}=-0.01\pm0.04$ 
\cite{Overseth}.
The equality $\Delta(\Xi^{0}_{0})=\Delta(\Xi^{-}_{-})=0$ is exact.
This is because there is only one final state isospin I=1 in $\Xi$ 
decays.  As a result, no final state phase difference is possible.

In order to obtain the predictions in various models more easily,
in the literature one continue to simplify the formula of CP
violating observables shown in Eqs.(15) and Eqs.(18). These formula
can be written as following \cite{Valencia,He}:
\begin{eqnarray}
\Delta(\Lambda^{0}_{-})&=&-2\Delta(\Lambda^{0}_{0})=\sqrt{2}
\frac{S_{33}}{S_{11}}sin(\delta_{3}-\delta_{1})sin(\theta^{S}_{3}-
\theta^{S}_{1}),\nonumber\\
A(\Lambda^{0}_{-})&=&A(\Lambda^{0}_{0})=-tan(\delta_{11}-\delta_{1})
sin(\theta^{P}_{1}-\theta^{S}_{1}),\nonumber\\
B(\Lambda^{0}_{-})&=&B(\Lambda^{0}_{0})=cot(\delta_{11}-\delta_{1})
sin(\theta^{P}_{1}-\theta^{S}_{1}),\nonumber\\
\Delta(\Xi^{-}_{-})&=&\Delta(\Xi^{0}_{0})=0,\nonumber\\
A(\Xi^{-}_{-})&=&A(\Xi^{0}_{0})=-tan(\delta_{21}-\delta_{2})
sin(\theta^{P}_{12}-\theta^{S}_{12}),\nonumber \\
B(\Xi^{-}_{-})&=&B(\Xi^{0}_{0})=cot(\delta_{21}-\delta_{2})
sin(\theta^{P}_{12}-\theta^{S}_{12}).\label{eqn:simp}
\end{eqnarray}
We have neglected the contributions
of $\Delta I=3/2$ in the CP violating observables except in $\Delta$.

The remaining task is to calculate weak interaction phases
$\theta_{i}$'s which in turn depend on the model for CP violation.
In following section we discuss the predictions of various models.

\section{Theoretical Predictions}
\ \ \ \
We summary the consequences of various CP violating models that has been 
discussed in the literature \cite{Valencia,He}in terms of tabular form, 
and their predictions are as following:

Table 1. Some models of CP violation in Hyperon decay.
\\
\begin{tabular}{|c|c|c|c|}
\hline
$\Lambda\;\mbox{decay}$&KM model&Weinberg Model& Left-Right Model\\ \hline
$\Delta(\Lambda^0_-)$& $<10^{-6}$&$-0.8\times 10^{-5}$&0\\
$A(\Lambda^0_-)$& $-(1\sim 5)\times 10^{-5}$& $-2.5\times 10^{-5}$&$-6\times
10^{-4}$\\
$B(\Lambda^0_-)$& $(0.6\sim 3)\times 10^{-3}$&
$1.6\times 10^{-3}$&$3.87\times 10^{-2}$\\ \hline
$\Xi\; \mbox{decay}$&&&\\ \hline
$\Delta(\Xi^-_-)$& 0&0&0\\
$A(\Xi^-_-)$&$(1\sim 10)\times 10^{-6}$&$3.31\times 10^{-5}$&$2.59\times
10^{-6}$\\
$B(\Xi^-_-)$&$-(1\sim 10)\times 10^{-3}$&$-3.76\times 10^{-2}$&$-2.94\times
10^{-3}$\\ \hline
\end{tabular}\\

In Left-Right Model for $\Lambda$ decay, 
  we have used the numerical results of Ref.\cite{DC}. For $\Xi$
decays, 
 the calculated results of Ref.\cite{LSW} with chiral perturbation have
been used. 
 From Table 1. we see that $\Delta$ is very small. 
Experimentally, it may be difficult to measure it. However, the prediction 
for the CP violating observable A is close to the region 
which will be probed by the E871 experiment with experimental sensitivity
$10^{-4}$ to $10^{-5}$ in $A_{asy}$.
\section {Summary}
\ \ \ \
The minimal SM of electroweak interaction is in very good
agreement with all experimental data so far, and there is no
evidence for any new particles or interactions below electroweak
scale. Although SM can give us CP violating effects
through three generation naturally, CP violation is still not considered
satisfactorily understood.

Here we have studied the phenomenology and introduced
some models to predict the CP odd observables $\Delta(\Lambda),
A(\Lambda)$ and $B(\Lambda)$ in hyperon system.
If we want to measure these direct CP asymmetry, final state
interactions are important.
From our analysis,
 we know that the predictions in SM are near the experimental sensitivity
in E871; and the predictions in WTHDM seem hard to reach
in the future. 
However, as we have shown in left-right model,
it may be the first candidate which can be measured
or tested in E871 proposal.

Given the crude estimates of the hadronic matrix elements involved
in various models,
all the numerical results should be viewed with caution. 
Nevertheless,
the results , in especial left-right model,
suggest that the search for CP violation in $A(\Lambda)$
at the $10^{-4}$ level of sensitivity that is expected for E871 is 
potentially very interesting and could reach. The suggestion is not only that
we can understand the mechanism of CP violation but also
that we may have the chance to find out new physics and
further understand the physics at electroweak scale. 
Besides SM, WTHDM
and left-right symmetric model, theoretically, we may also
consider the contribution of the other models such as supersymmetric
model or unified theory. Whatever we bulid the models, we still
hope that the experiment could tell us what CP violation is. We need
more experimental data to get the nontrivial values for  
neutron electric dipole moment (NEDM) and $\epsilon'/\epsilon$.

We conclude that it is possible for E871 to observe a CP violating signal 
at the $10^{-4}$ level.

\section*{ACKNOWLEDGEMENTS}
This work is supported in part by the National Science Council
of Republic of China under grant No. 85-2112-M-007-032 and -029.


\end{document}